\documentclass[10pt]{iopart}
\usepackage{graphicx}
\usepackage{gensymb}
\expandafter\let\csname equation*\endcsname\relax
\expandafter\let\csname endequation*\endcsname\relax
\usepackage{mathptmx}
\usepackage{verbatim}
\usepackage{amsmath}
\usepackage{amssymb}
\usepackage{amsfonts}
\usepackage{mathrsfs}
\usepackage{amsmath} 
\usepackage{color}
\usepackage{bm} 
\usepackage{xspace}
\usepackage{hyperref}
\usepackage{multirow}
\usepackage{color}
\usepackage{float}
\usepackage{bbold}
\usepackage{tabu}
\usepackage{textcomp}
\usepackage{dcolumn}
\usepackage{diagbox} 

\begin{document}
\title[]{Commensurate and incommensurate double moire interference in graphene encapsulated by hexagonal boron nitride}
\author{N. Leconte$^1$, J. Jung$^2$}
\address{$^1$ Department of Physics, University of Seoul}
\ead{jeil.jung@uos.ac.kr}
\begin{abstract}
Interference of double moire patterns of graphene (G) encapsulated by hexagonal boron nitride (BN) 
can alter the electronic structure features near the primary/secondary Dirac points and the electron-hole 
symmetry introduced by a single G/BN moire pattern depending on the relative stacking arrangements 
of the top/bottom BN layers.
We show that strong interference effects are found in nearly aligned BN/G/BN and BN/G/NB and 
obtain the evolution of the associated density of states as a function of moire superlattice twist angles. 
For equal moire periods and commensurate patterns with $\Delta \phi = 0^{\circ}$ modulo $60^{\circ}$ 
angle differences the patterns can add up constructively leading to large pseudogaps of about 
$\sim 0.5$ eV on the hole side or cancel out destructively depending on their relative sliding, e.g. 
partially recovering electron-hole symmetry.
The electronic structure of moire quasicrystals for $\Delta \phi =30^{\circ}$ differences reveal double
moire features in the density of states with almost isolated van Hove singularities where we can expect strong correlations. 
\end{abstract}
\maketitle
\ioptwocol

\section{Introduction}
\label{introdSect}

Research on vertical heterostructures of atomically thin two-dimensional van der Waals (vdW) 
materials has been a booming field~\cite{Koma_1992, Koma_1999, Geim:hf}
thanks to the new opportunities for tailoring artificial materials with novel electronic properties,
for which magic angle twisted bilayer graphene~\cite{Bistritzer_2011} or trilayer graphene on hBN~\cite{Chen_2019,Chen_2019b}
have emerged as prototypical systems where signatures of Mott insulating phases and superconductivity have been observed.
In particular, the hexagonal boron nitride (BN)~\cite{Dean_2010} is an excellent substrate 
material that preserves the properties of pristine graphene (G) 
and introduces moire super-lattice features 
at experimentally accessible magnetic fields and gate voltages when they are nearly aligned~\cite{Yankowitz_2012, Ponomarenko_2013, Hunt_2013, Dean_2013, PhysRevB.87.245408, Jung_2014, Jung_2015}. 
Experiments in G/BN systems can make use of an additional capping BN dielectric film
that further screens the system from extrinsic disorder and improves the device mobilities~\cite{Mayorov2011, Taychatanapat2013}. 
These are often deposited at wide twist angles to avoid potential interference with the moire pattern of the BN substrate.
With increasing precision in the rotation angle control of 2D materials~\cite{Frisenda_2018,Kim_2017} it is desirable to understand what would be the combined 
effects where both encapsulated layers are nearly aligned with the graphene layer.
Recent experiments  of graphene encapsulated by nearly aligned boron nitride sheets 
indicate enhancement of primary Dirac point band gaps and additional density of state peaks near the secondary Dirac 
point~\cite{Finney2019,NLWang} that can be expected from the momentum conservation conditions for the electrons
traveling in a double moire system~\cite{1910.00345}.
The superposition of moire patterns has also been considered as a tool to characterize multilayer 
systems via atomic moir\'e interferometry~\cite{PhysRevB.81.125427} 
and traces of this interference on the electronic structure have also been observed at finite magnetic fields~\cite{NLWang}.
Other terminology used include super-moires (SM)~\cite{1910.00345, Wang2019} in graphene encapsulated by boron nitride,
and equivalently moire of moires~\cite{1912.03375} in twisted trilayer graphene.

In this paper we investigate the effects of superposing two moire patterns in the parameter space of twist angles 
$( \theta_1, \, \theta_2 )$ and sliding vector magnitudes
$(  \tau_1,  \, \tau_2 )$ of the BN sheets at each interface that encapsulate the reference graphene sheet.
We pay particular attention to the double moire structures that have not been addressed in earlier literature, 
namely the different commensurate and quasi-crystal limits 
where the double moire features show up in a special manner. 
The resulting moire patterns will have a relative rotation of $\Delta \phi = \phi_1 - \phi_2$ and for equal moire periods 
we can achieve commensurate patterns for $\Delta \phi = 60^{\circ}$ angle differences while moire quasicrystals 
are expected for $\Delta \phi =30^{\circ}$ differences.
The interference between the secondary Dirac point features for nearly aligned moire patterns can add up constructively or destructively, 
depending on both the relative sliding between bottom and top hBN layers as well as the respective orientation of the top layer with 
respect to the bottom layer ($\Delta \theta \sim 0^{\circ}$ or $\Delta \theta \sim 60^{\circ}$) at energies where families of 
super-moire features converge. 
We also observe that in the incommensurate regime, the relative strength of super-moire features are only weakly dependent 
on the orientation and the periods of the constituent moire patterns, while 
strongest features happen close to charge neutrality and near the secondary Dirac point
on the hole side where significant suppressions in the density of states take place.

\section{Electronic structure of aligned BN/G/BN and BN/G/NB configurations} 
\begin{figure*}[t]
\begin{center}
\includegraphics[width=2\columnwidth]{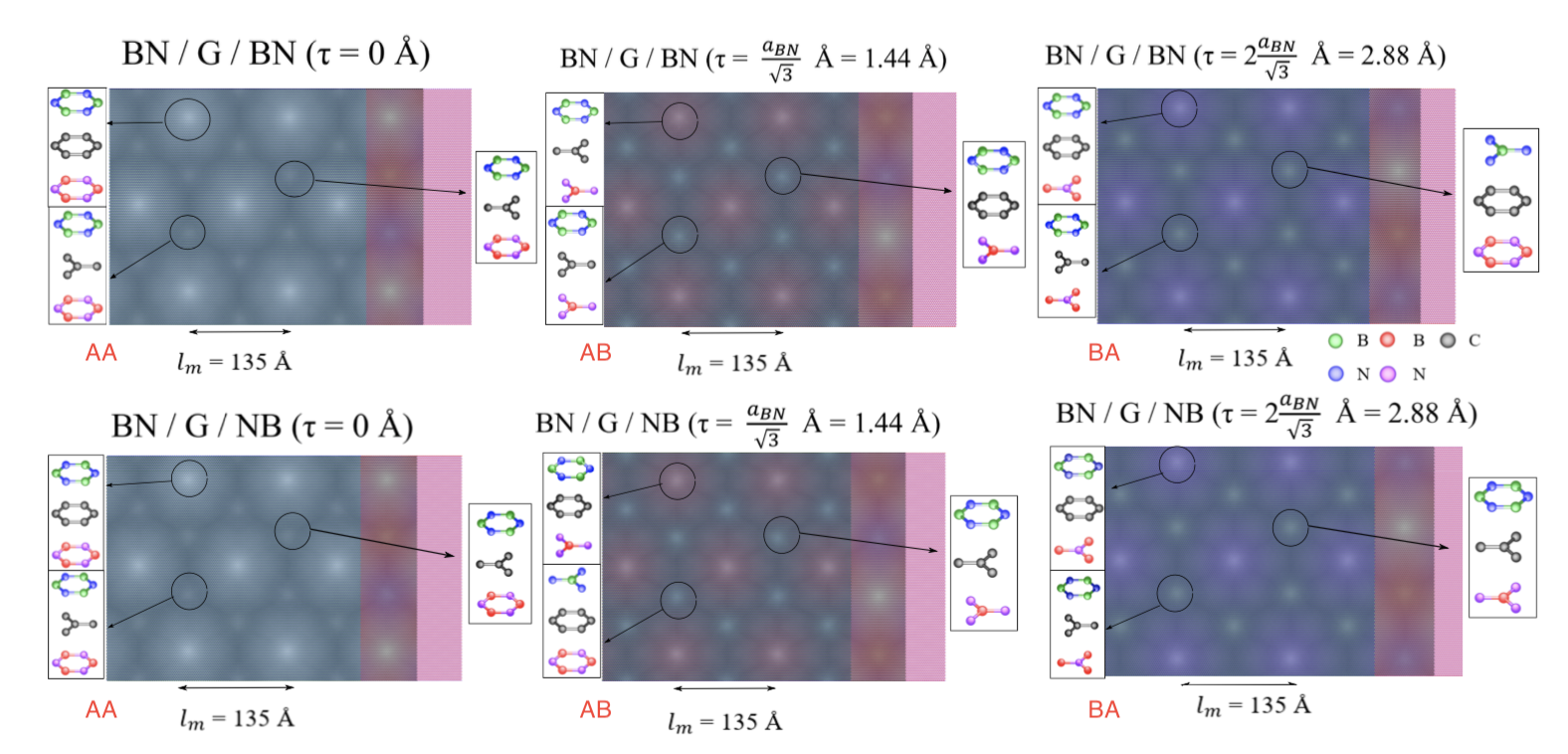}
\caption{\label{figure1} Moire patterns for a select number of slidings between the bottom and top 
layer of hBN including local stacking sketches representing the two investigated systems, 
BN/G/BN and BN/G/NB ($60^{\circ}$ rotation) with carbon atoms (grey), 
boron atoms (green or red) and nitrogen atoms (pink or blue). 
For BN/G/BN and BN/G/NB aligned systems.
The magnitude of $\tau$ represents the relative sliding along the y-direction of the top BN layer with respect to the bottom layer.
}
\end{center}
\end{figure*}

\begin{figure}[t]
\begin{center}
\includegraphics[width=1\columnwidth]{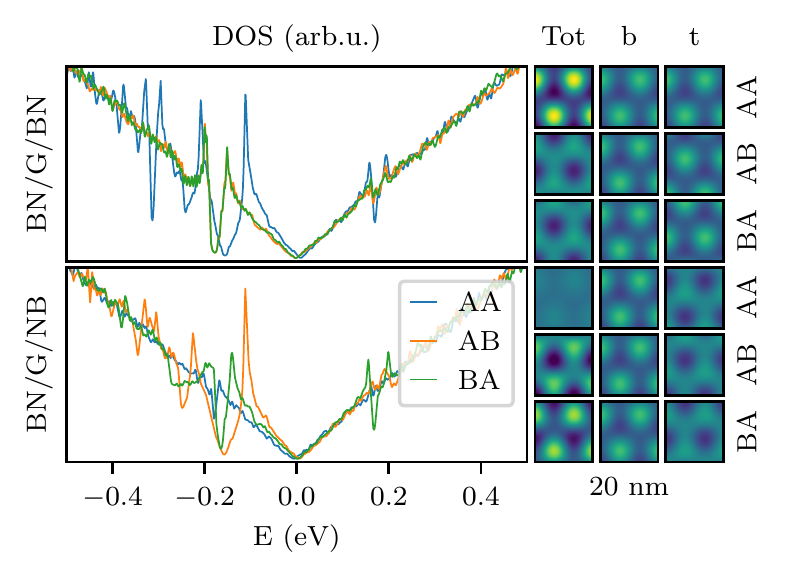}
\caption{\label{figure2_dos} 
(Color online) 
The density of states plots and $H_0 + H_z$ moire pattern contributions in Eq.~\ref{moire}
corresponding to the three different stacking arrangements between the bottom and top encapsulating BN layers
that we label as AA, AB, BA where we fix the bottom layer moire pattern and slide the top
layer moire along the $y$-axis by 0, $\lambda / \sqrt{3}$, and $2 \lambda / \sqrt{3}$.
We consider perfect aligment of both encapsulating BN layers 
corresponding to BN/G/BN and 
BN/G/NB alignments, referred to hereafter as $C_0^{(\prime)}$, respectively.
}
\end{center}
\end{figure}

Let us consider graphene sandwiched between two BN layers
where the graphene layer is the fixed reference frame and the twist angles $\theta_1$ and $\theta_2$ 
refer to the bottom and top BN layers respectively. 
The simplest commensurate double moire structures can be formed when
$\theta_1 = \theta_2 = 0^{\circ}$ where the two BN layers have the same orientations which we label as BN/G/BN, 
or $\theta_1 = 0^{\circ}$ and $\theta_2 = 60^{\circ}$ configurations where the top BN layer 
has a different alignment which we label as BN/G/NB, see Fig.~\ref{figure1}.
We define the sliding vector along the $y$-axis ${\bm \tau} = (0,\tau) = {\bm \tau}_2 - {\bm \tau}_1$ to represent
the difference between the top and bottom BN layers with respect to the central graphene reference system.
The effective Hamiltonian of graphene subject to the moire potentials stemming from 
both BN layers can be written as
\begin{equation}
H  =  H_{\rm G} + H_{M}^b + H_{M}^t    
\label{moire} 
\end{equation}
where the Hamiltonian of graphene is modeled with a finite gap of $2\Delta_0$
that appears due to alignment with BN
\begin{equation}
 H_{\rm G}    = \hbar \upsilon_{\rm F} {\bm p} \cdot {\bm \sigma} \tau_0 + \Delta_0 \sigma_3 \tau_3 
\end{equation}
and we use the Pauli matrices $\sigma_{i}$ and $\tau_i$ that operate on the sublattice 
and valley pseudospins respectively. 
Since the primary focus of our work is for the interference effects in double moire patterns
for the analysis presented in this work we set the primary Dirac point gap $2\Delta_0$ to zero 
but they should be accounted for explicitly in a theory that intends to resolve the band gaps near charge neutrality. 
The moire patterns for the top and bottom interfaces are added to account for the double moire
\begin{equation}
 H^{l}_{M} = H_{0}^{l} ({\bm r}) \sigma_0 \tau_0 + H_z^{l} ({\bm r}) \sigma_3 \tau_3 + {\bm H}_{xy}^{l} ({\bm r}) \cdot {\bm \sigma} \tau_3  
\label{Htot}
\end{equation}
where the indices can take $l = \pm 1$ values depending on the G/BN $(+)$ or G/NB $(-)$ orientations
irrespective to whether the BN layer is above or below the graphene layer.
The moire pseudospin components are
\begin{equation}
H_0^l \left({\bm r}\right)= 2 C_0 \Re e \left[ f^{l}\left({\bm r}\right) e^{i \phi_0}\right],
\label{H0eq}
\end{equation}
\begin{equation}
H_z^l \left({\bm r}\right) = l 2 C_z  \Re e \left[ f^{l}\left({\bm r}\right) e^{i \phi_z}\right],
\label{Hzeq}
\end{equation}
for the diagonal terms with  $f^{l}({\bm r}) = \sum_{m}\,\exp(i l {\bm G}_m\cdot {\bm d}({\bm r})) \, \,(1+  (-1)^m)/2$
which is complex conjugated when the sign of $l$ is reversed.
We use the  reciprocal lattice vectors ${\bm G}_m$ of the reference frame lattice and
the local stacking function 
\begin{equation}
{\rm d}({\bm r}) = (\alpha {\cal R}(\theta) - 1) {\bm r}
\end{equation}
where $\alpha = a / a_{\rm ref} = 1 + \varepsilon$ is the scaling ratio with respect to the reference lattice 
and ${\cal R}(\theta)$ is the rotation operator acting on an arbitrary point in real space at lattice $\bm r$.
The off-diagonal component is given by
\begin{equation}
\label{Hxyeq}
{\bm H}^l_{xy} ({\bm r})   = 2C_{xy} \cos(\phi) \, \hat{\bm z} \times {\bm \nabla}\,\Re e[e^{i\phi_{xy}}f^{l}({\bm r})]
\end{equation}
where we use the vector notation to distinguish the real and imaginary parts.
The moire pattern rotation angle $\phi$ for small $\theta$ satisfies
$\cos (\phi) \approx \varepsilon/(\varepsilon^2 + \theta^2)^{1/2}$~\cite{PhysRevB.96.085442}.
The tight-binding (TB) model in real space is obtained by mapping the continuum moire patterns
into real-space as introduced in Ref.~\cite{Leconte_2016}.
The sublattice diagonal terms of the TB Hamiltonian can be mapped
in a straightforward manner as site potential energies from Eqs.~(\ref{H0eq}) and~(\ref{Hzeq}).
For the off-diagonal term we note that ${\bm H}_{AB} = {\bm H}_{xy}^{*}$
following the definition of the Pauli matrices,
and using the sublattice definitions of Refs.~\cite{PhysRevB.96.085442, Leconte_2016} we have
\begin{equation}
{\bm H}_{\rm AB} = 
\delta_1 - \frac{\delta_2 + \delta_3}{2} + i \frac{\sqrt{3}}{2} (\delta_3 - \delta_2),
\label{HAB3}
\end{equation}
where the $\delta_i$ correction terms to the pristine graphene hopping of $t_0 = -3$ eV capture the unequal hopping amplitude of the electrons from A to B sites
due to the moire pattern of strains and they are given by 
\begin{equation}
\delta_1 =  \frac{2}{3} \Re e({\bm H}_{\rm AB}), \quad \delta_{2,3} 
=  \frac{- \Re e({\bm H}_{\rm AB}) \pm \sqrt{3} \, \Im m({\bm H}_{\rm AB})}{3}. 
\label{delta1}
\end{equation}
where $\Re e({\bm H}_{\rm AB})$ and $\Im m({\bm H}_{\rm AB})$ are evaluated at the 
Brillouin-zone corner of graphene
${\bm K} = ( 4 \pi / 3a_{\rm G},0)$.
The parameters defining our model correspond to the non-relaxed G/BN system and they are given by
~\cite{PhysRevB.96.085442}
\begin{align}
& C_0 = 10.13 \text{meV},  &\phi_0 = 146.53  \degree, \nonumber \\
& C_z = 9.01 \text{meV},   &\phi_z = 68.43 \degree, \nonumber \\ 
& C_{xy} = 11.34 \text{meV},  & \phi_{xy} = -109.6 \degree.
\label{params}
\end{align}
that are equivalent to those of Ref.~\cite{Jung_2014} but have been updated to use positive magnitudes
for the coefficients and the redefined phase $\phi_{xy} = \pi /6 - \phi_{AB}$
following the conventions in Ref.~\cite{PhysRevB.96.085442}.
The density of states are calculated using the Lanczos recursion method~\cite{arXivFan, Leconte_2011} using 10000 recursion steps on about 80 million atom systems for the data in Figs.~\ref{figure2_dos}, \ref{CommensurateSensitivity} and~\ref{strongestfeatures} and 4000 recursion steps on about 20 million atoms for the DOS maps in Figs.~\ref{PRLFig1}. The energy broadening in the Lanczos method is increased from its arbitrarily small value in the previous maps to about 1.5~meV for the DOS curves in Figs.~\ref{figure2_dos} and \ref{CommensurateSensitivity} to damp numerical oscillations.

For the aligned and commensurate moire geometries, referred to as $C_0^{(\prime)}$ hereon, the relative sliding between the top and bottom BN layers can 
give rise to very different electronic structures, see Fig.~\ref{figure2_dos} for the plot of the different density of states. 
We can for instance recover a certain electron-hole symmetry due to destructive interference effects 
for BN/G/NB in AA-stacked arrangement, as rationalized by the vanishing features in the corresponding 
$H_0 + H_z$ moire contribution maps, or find strong increases in the DOS dip due to 
constructive interference effects as compared to BN/G for BN/G/BN in AA arrangement
and in agreement with recent experimental observations on such systems~\cite{Finney2019,Wang2019}.
%

\section{Commensurate and quasi-crystalline double moire patterns}

In our simulations, we use the effective model for the moire pattern potentials
consistent with the conventions for the reference frame and relative lattice constants
outlined in Ref.~\cite{Jung_2014} where
%
the moire pattern angle $\phi$ for a layer rotated by $\theta$ from a reference frame 
is given by 
\begin{equation}
\phi = \tan^{-1}\left(\frac{\alpha \sin{\theta}}{\alpha\cos{\theta}-1}\right)
\label{phiEq}
\end{equation}
using the scaling parameter $\alpha = 1+\varepsilon$ accounts for 
the lattice constant mismatch $\varepsilon = (a - a_{\rm ref} )/ a_{\rm ref}$ with respect to a reference frame 
lattice constant $a_{\rm ref}$, where the minor differences with respect to Ref.~\cite{Yankowitz_2012} are related 
with the conventions in the choice of the reference frames.  
When one BN layer is used as a reference frame lattice an approximate fractional 
value $\varepsilon = (  a_{\rm G} - a_{\rm BN}  )/ a_{\rm BN} = - 1/55$
can to account for the $\sim 1.7\%$ reduction of graphene's lattice constant with respect to that of BN
while facilitates constructing commensurate moire supercells.
The period of the resulting moire pattern is given by $\lambda = a_{\rm ref} / [\varepsilon^2 +  ( 1 + \varepsilon)(2 - 2 \cos \theta)]^{1/2}$.

\begin{figure}[t]
\label{smschematic}
\begin{center}
\includegraphics[width=8cm]{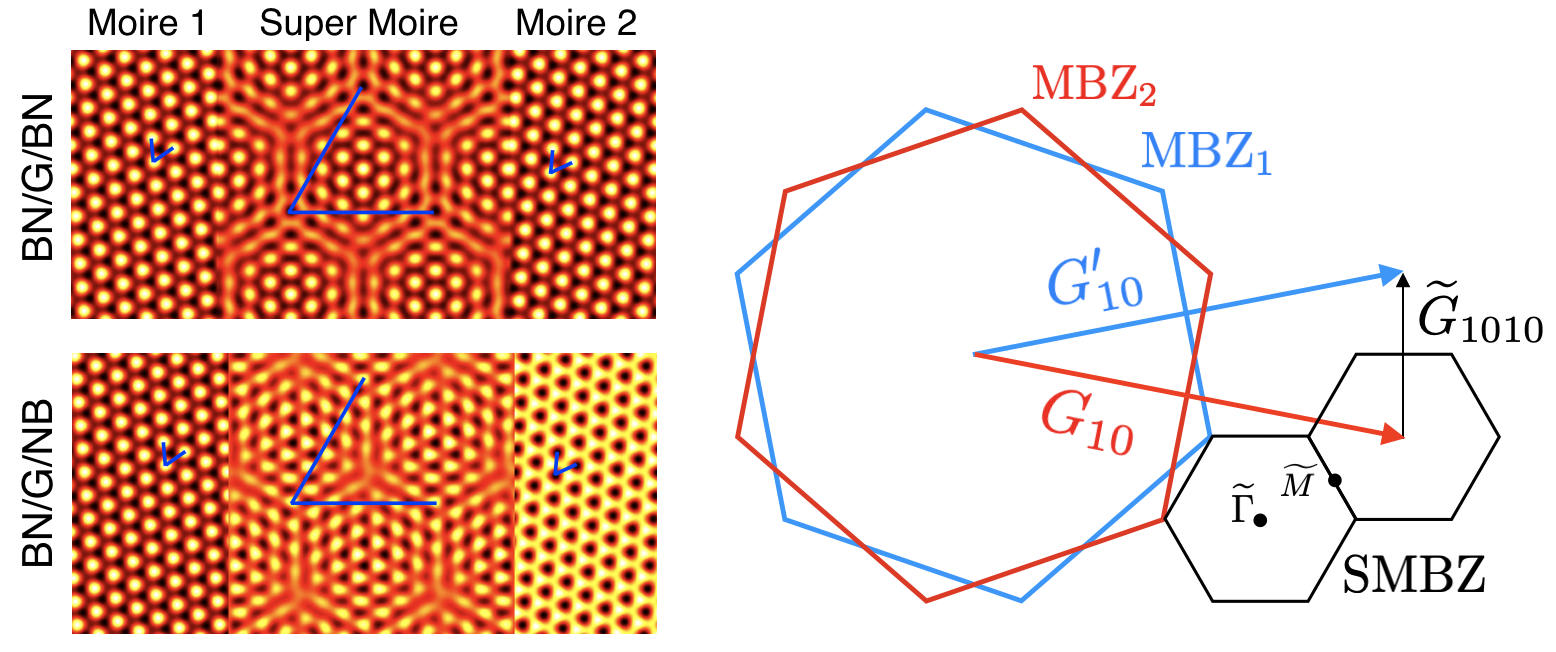}
\caption{
(color online) {\em Left panel:} 
Superposition of two equal period moire patterns, namely moire 1 and moire 2,  that are twisted by an 
angle of $\Delta \phi$ and the supermoire (SM) pattern that arises by their superposition. 
We use opposite contrasts for moire 2 depending on the BN or NB overlayer. 
{\em Right panel:} Schematic representation of the two moire Brillouin zones (MBZ) that are rotated with respect 
to each other and the SM Brillouin zone (SMBZ) resulting from their interference through Eq.~(\ref{mconserv}). 
The SM features are expected near the $\widetilde{M}$ points between the SMBZ at energies
given by Eq.~(\ref{moireEnergy}) assuming a linear dispersion of the bands.
}
\end{center}
\end{figure}

In this section we consider supermoires (SM) arising from double moires 
for which we can use the same formula as in the ordinary moire patterns stemming from two rotated lattices
in Eq.~(\ref{phiEq}) by substituting $a_i \rightarrow \lambda_i $ and $\theta \rightarrow \Delta \phi$ 
to obtain the supermoire lengths $\widetilde{\lambda}_{\rm SM}$ and twist angles $\widetilde{\phi}_{\rm SM}$.
The electrons in a double moire that gives rise to a SM obey momentum conservation rules that are similar to that of electrons in incommensurable double lattices within a moire superlattice. The reciprocal lattice of the SM is given by
%
%
\begin{equation}
\widetilde{ \bm G}_{pqrs} = {\bm G}^{\prime}_{pq} - {\bm G}_{rs}
\label{mconserv}
\end{equation}
where ${\bm G}_{pq}^{\prime} = p {\bm G}^{\prime}_{10} + q {\bm G}^{\prime}_{01}$ 
and ${\bm G}_{rs} = r {\bm G}_{10} + s {\bm G}_{01}$ are the reciprocal lattice vectors
associated to each one of the two moire patterns~\cite{ster}. 
We note that multiple SM reciprocal lattices are possible 
but we will restrict our attention to $ p,\, q, \, r, s = 0,1$ values of the moire lattices.
In Fig.~\ref{smschematic} we show a schematic representation of a SM formed by 
two equal period moire potentials twisted in opposite angles.

\begin{figure*}[t]
\begin{center}
\includegraphics[width=2\columnwidth]{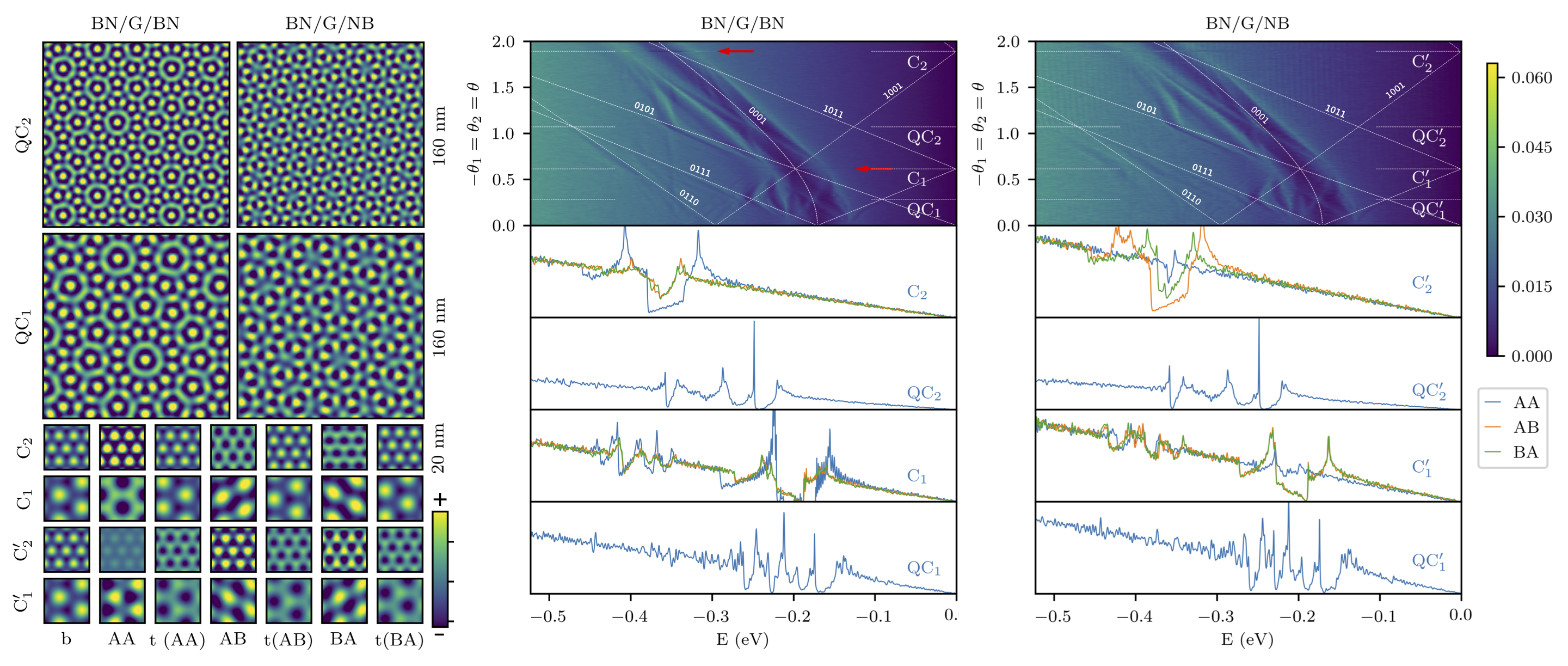}
\caption{
\label{PRLFig1} 
(color online) {\em Left panel:} Onsite energy maps ($H_0 + H_z$) for QC$^{(\prime)}_1$, QC$^{(\prime)}_2$, 
C$^{(\prime)}_1$ and C$^{(\prime)}_2$, 
for BN/G/BN and BN/G/NB aligned systems the latter referenced by primes. 
The $b$ and $t$ subscripts refer to the bottom and top interface moire contributions between the encapsulating BN layers. 
We label by (AA, AB and BA) the stacking geometries for commensurate geometries where we fix the bottom layer moire 
and slide along the $y$-axis the top layer moire by 0, $\lambda / \sqrt{3}$, and $2 \lambda / \sqrt{3}$.
{\em Middle and right panels:} Density of states for combinations of $\theta_1 = -\theta_2$ for BN/G/BN (middle) 
and BN/G/NB (right) as a function of twist angles. 
We show DOS plots for specific twist angles corresponding to commensurate cases ($C_1$ and $C_2$)
whose electronic structure depends on the sliding of the top BN layer, 
and quasi-crystal structures $QC_1$ and $QC_2$ insensitive to sliding 
where the moire patterns make an angle of $\Delta \phi = 30^{\circ}$.
Electronic structure features on the electron side are smaller and not shown here. 
The fine white dashed lines are plotted to guide the eye and are obtained for $p$, $q$, $r$, $s < 2$
from Eq.~\ref{moireEnergy}) for the electronic structure features at SM zone boundaries.
Actual differences between these lines and the calculated DOS are due to the deviation of the moire bands from the
Dirac model bands and they can be fitted by modifying the effective Fermi velocity or improving the 
reference energy band model. 
}
\end{center}
\end{figure*}
The SM features are expected at energies $\pm E({\bm k}_{\widetilde{M}})$ in the SMBZ
boundaries at momenta ${\bm k}_{\widetilde{M}, \, pqrs} = \widetilde{\bm G}_{pqrs}/2$.
Assuming a linear Dirac dispersion these energies are 
\begin{equation}
%
E_{pqrs} =  \upsilon_{\rm F} \frac{ \hbar \widetilde{G}_{pqrs}}{2}
\label{moireEnergy}
\end{equation}
where we use the notation $\widetilde{G}_{pqrs}  = \left| \widetilde{\bm G}_{pqrs} \right|$ for each set 
of $p, q, r, s$ values which reduces to the form proposed in Refs.~\cite{Wang2019,ster}
at energies $E_D = \pm 2 \pi \upsilon_{\rm F} / (\sqrt{3} \widetilde{\lambda}_{{\rm SM}, \,pqrs})$ for a SM period given by 
$\widetilde{\lambda}_{{\rm SM}, \,pqrs} = 4\pi/ (\sqrt{3} \widetilde{G}_{pqrs} )$.

We begin by considering equal period moire patterns for the bottom and top layers
for variable $\theta = \theta_1 = -\theta_2$ and we illustrate in Fig.~\ref{PRLFig1} 
the colormap for hole density of states (DOS) for small twist angles up to $\theta = 2\degree$. 
On top of it we also plot as a guide to the eye the dashed white lines 
the energies at the $\widetilde{M}$ point of the SMBZ assuming a 
Dirac cone energy dispersion given in Eq.~(\ref{moireEnergy})
for SM reciprocal lattice vectors corresponding to $p$, $q$, $r$, $s$ = 0, 1. 
These energies show good overall agreement with features in the DOS obtained from explicit 
electronic structure calculations while deviations thereof 
can be attributed to the moire superlattice band features in G/BN
that distort the Dirac Hamiltonian.
%
%
%
%
%
%
%
%
These select angles for the C and QC solutions obtained from Eq.~(\ref{phiEq})  are
$\theta_{C_1} = 0.61^{\circ}$,  
$\theta_{C_2} = 1.89^{\circ}$,  
$\theta_{QC_1} = 0.28^{\circ}$, and  
$\theta_{QC_2} = 1.07^{\circ}$.  
We use the prime symbols to denote a rotation by $60^{\circ}$ of the top moire system 
that switches the respective positions of top layer B(oron) and N(itrogen) atoms.
From Fig.~\ref{PRLFig1} we observe that for those select twist angles the multiple SM features merge to potentially enhance 
the electronic structure features.
As an illustration of the double moire patterns for those select angles 
we plot the real-space map of the $H_0 + H_z$ terms defined in Eqs.~(\ref{H0eq}) 
and (\ref{Hzeq}) where we can observe commensurate triangular double moire patterns 
with $\Delta \phi = 0^{\circ}$ for C$_1$, C$_2$ cases,
and dodecahedral quasicrystal patterns with $\Delta \phi =30^{\circ}$ double moire patterns 
for QC$_1$, QC$_2$ cases. 

\begin{figure}[t]
\begin{center}
\includegraphics[width=1\columnwidth]{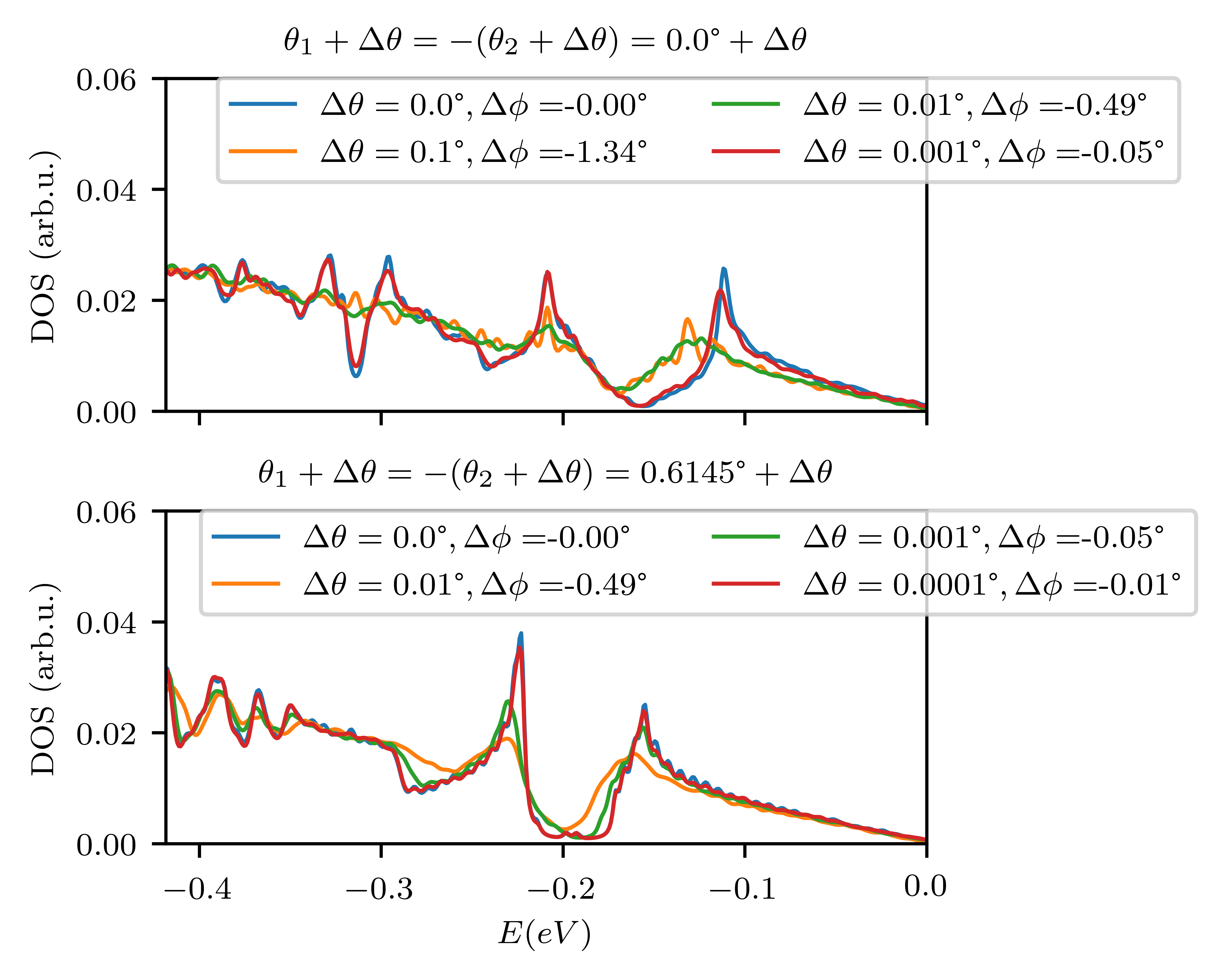}
\caption{\label{CommensurateSensitivity} 
(Color online)  The strong commensurate features here for two selected cases: 
C$_0$ from Fig.~\ref{figure2_dos} and C$_1$ from Fig.~\ref{PRLFig1} in the AA stacking 
arrangements show a high sensitivity of features to misalignment because 
already when both layers are rotated by more than $\theta_{1(2)} \pm 0.01 \degree$ respectively, 
the strong interference features are drastically modified. 
We expect that the sensitivity to twist angle $\theta$ will be weaker in practice provided 
that the tendency for the moire twist angles $\phi$ to mutually align through local strain fields
is strong enough in experiments and will determine the actual possibility for realization 
and control of nearly commensurate double moire systems. 
}
\end{center}
\end{figure}

The solutions for commensurate double moires C$_1^{(\prime)}$, C$_2^{(\prime)}$ show up as 
singular discontinuous points in the phase space of twist angles that we marked with red arrows in the middle panel of Fig.~\ref{PRLFig1} and these are points where the SM features from different moire reciprocal lattice vectors cross each other at those specific energies, signaling potentially stronger interference of the moire features. 
The solutions are found to depend strongly on the specific sliding of the top BN layer moire pattern relative to bottom where we allowed the top layer moire to slide along the $y$-axis by 0, $\lambda/\sqrt{3}$ and $2 \lambda/\sqrt{3}$ to for AA, AB and BA high symmetry stacking configurations.
While the commensurate features in Fig.~\ref{figure2_dos} and~\ref{PRLFig1} show striking double moire interference
the discontinuities indicated by the red arrows in Fig.~\ref{PRLFig1} also suggest they are quite sensitive to small departures from perfect alignment. We illustrate this point in Fig.~\ref{CommensurateSensitivity} where we show the DOS for 
the simple commensurate double moire consisting of two aligned BN layers with $\theta = 0^{\circ}$ (C$_0$) and the C$_1$ case. 
We see that in both panels, the strong features stemming from constructive interference, 
including secondary gaps at around $-0.16$~eV in the top panel and $-0.2$~eV in the bottom panel, 
have already vanished for twist angles of the order of $\Delta \theta = 0.01\degree$ because these 
twists imply a much larger mismatch of the order of $\Delta \phi = 0.5\degree$ between the 
orientations of the respective moire patterns. Yet, experimental observations of practically aligned commensurate double moire systems~\cite{Finney2019} suggest that the moire systems might energetically lock into maximum alignment thus providing support that commensurate double 
moire patterns are within experimental reach.
Strongest SM features are expected for the AA stacked C$_1$ case where we find a remarkable suppression in the density of states
near the secondary Dirac point showing a rather large gap-like structure $\sim$0.5~eV. 
This value is an order of magnitude larger than observed secondary Dirac point gaps of $\sim$12~meV in a 
single G/BN interface resolved in recent experiments~\cite{doi:10.1021/acs.nanolett.8b03423} and would show unequivocally in experiments. 
Although these features are partially suppressed when the double moire pattern is modified 
to the AB or BA stacking, there is still a remarkable 
suppression in the electronic density of states around the secondary Dirac point energies
far greater than that produced by a single moire pattern. 
We notice a progressive suppression of the moire features in the C$_1^{\prime}$ configuration
where a similar gap-like feature survives for AB and BA stacking but largely disappears
for AA stacking, due to the mutual cancellation of the moire patterns effects between the top and bottom layer moires. 
In the C$_2^{\prime}$ cases we see significant secondary Dirac point features 
for greater twist angles at energies of $\sim0.36$~eV
which should be accessible through ionic liquid gating techniques
but are outside reach using ordinary gating techniques based on 
low permittivity dielectric barrier materials.

The QC$_1$ and QC$_2$ moire quasicrystals are formed for specific twist angles where
$\Delta \phi = 30^{\circ}$. The associated DOS maps show clear traces of 
SM features stemming from moire interference effects with a large number of spikes
that are partially smoothened by the broadening intrinsic to the calculations. 
Multiple split DOS spikes are result from the interference effects and the features are strongest 
in the vicinity of the energies where the G/BN secondary Dirac point dips exist on the hole side. 
The major van Hove singularity peaks near $\sim -0.18$~eV and 
near $\sim -0.26$~eV for QC$_1$ are surrounded by suppressions 
in the DOS leading to a practically isolated DOS peaks
making in principle the QC double moire systems a promising platform for 
finding Coulomb interaction driven physics.
The QC$_2$ shows a well defined prominent van Hove singularity peak 
that is located farther away from charge neutrality than in QC$_1$ but yet
at carrier densities that should be accessible in experiments.  
Differences between BN/G/BN and BN/G/NB encapsulations are rather weak in 
the DOS maps although their real space LDOS maps may show contrasting behaviors.
Further studies on the properties of these moire quasicrystals will be presented elsewhere.

\begin{figure}[t]
\begin{center}
\includegraphics[width=1\columnwidth]{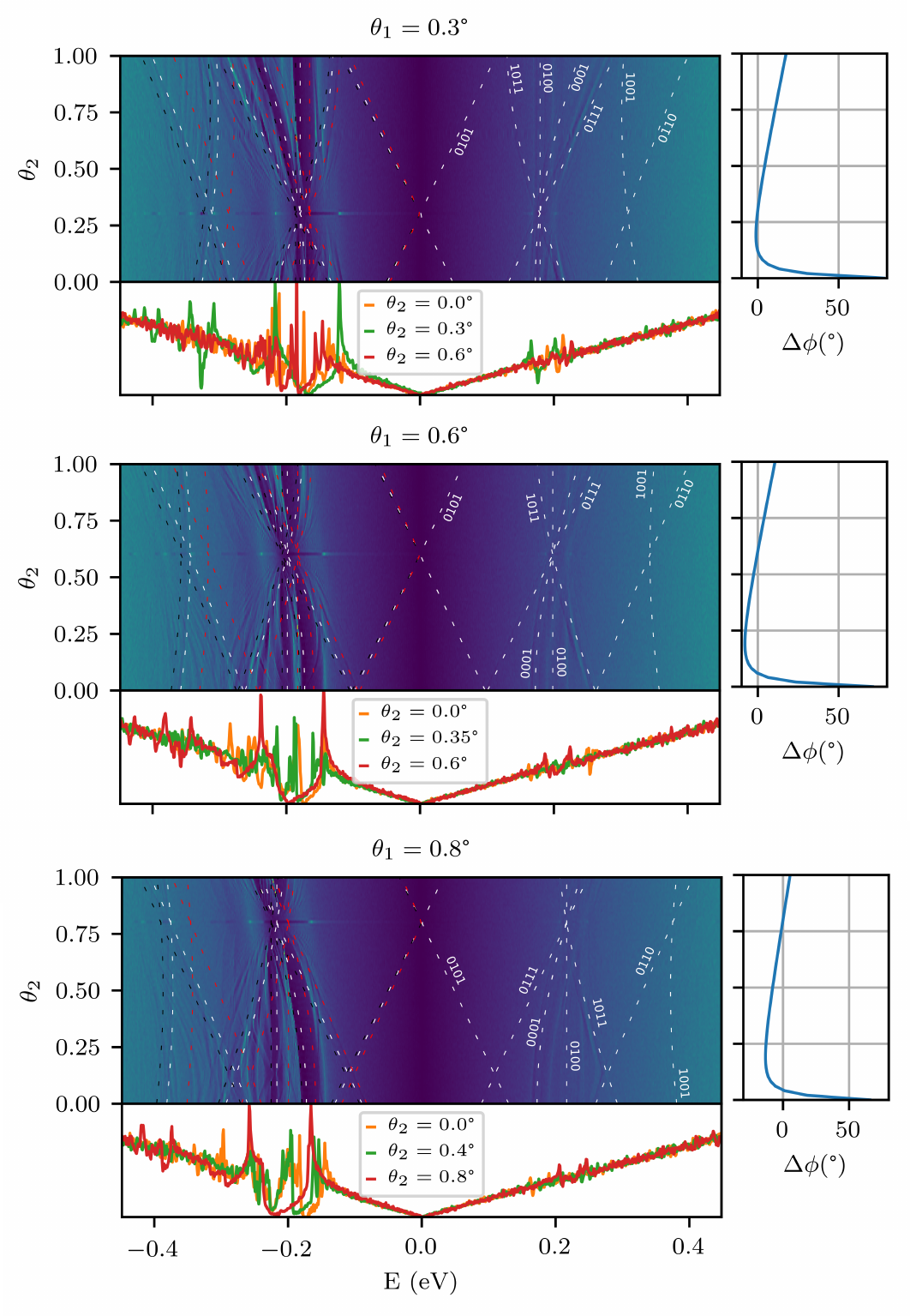}
\caption{\label{strongestfeatures} 
(Color online)  Density of states colormaps for fixed bottom layer with $\theta_1 = 0.3 \degree$ (top panel), 
$0.6 \degree$ (middle panel) and $0.8 \degree$ (bottom panel) with rotation $\theta_2$ of the top layer 
ranging from $0$ to $1\degree$ for the BN/G/BN system. 
On the right hand side of each panel we show the orientation angle differences $\Delta \phi$ between the 
constituent moire patterns showing stronger sensitivity for small $\theta$ values. 
At the bottom of each panel, we show the line cuts of the DOS for select values of $\theta_2$,
including the  $\theta_1 = \theta_2$ and AA commensurate cases where we see enhancement
of their features at select energies due to the merger of multiple supermoire features into a few.
%
%
We label the $pqrs$-tuplet [see Eq.~(\ref{moireEnergy})] labels on the electron side while the symmetric lines for holes
are ommitted. The black, white and red dashed lines correspond to 
$\upsilon_{\rm F} = 1.05 \upsilon_{{\rm F}_0}$, $1.01 \upsilon_{{\rm F}_0}$ and 
$0.93 \upsilon_{{\rm F}_0}$ respectively, where $\upsilon_{{\rm F}_0}  = \sqrt{3}\, t_0 \, a_G / (2 \hbar)$ and $t_0 = -3.0$ eV.
The renormalization of the Fermi velocities and captures the deviation from the Dirac bands.
Fermi velocity renormalization (black dashed lines match better around $-0.35$ eV, white lines around $-0.19$~eV, 
and red lines around $-0.12$ eV), $\upsilon_{\rm F}$ drops more strongly when the range of superlattice features is larger, 
e.g., in the middle panel, for $\theta_1 = 0.6\degree$ and $\theta_2 = 0\degree$, between $-0.25$ and $-0.12$ eV, 
than when $\theta_2 = \theta_1 = 0.6 \degree$, leading to further bending of the geometric lines.}
\end{center}
\end{figure}

\begin{figure}[t]
\begin{center}
\includegraphics[width=1\columnwidth]{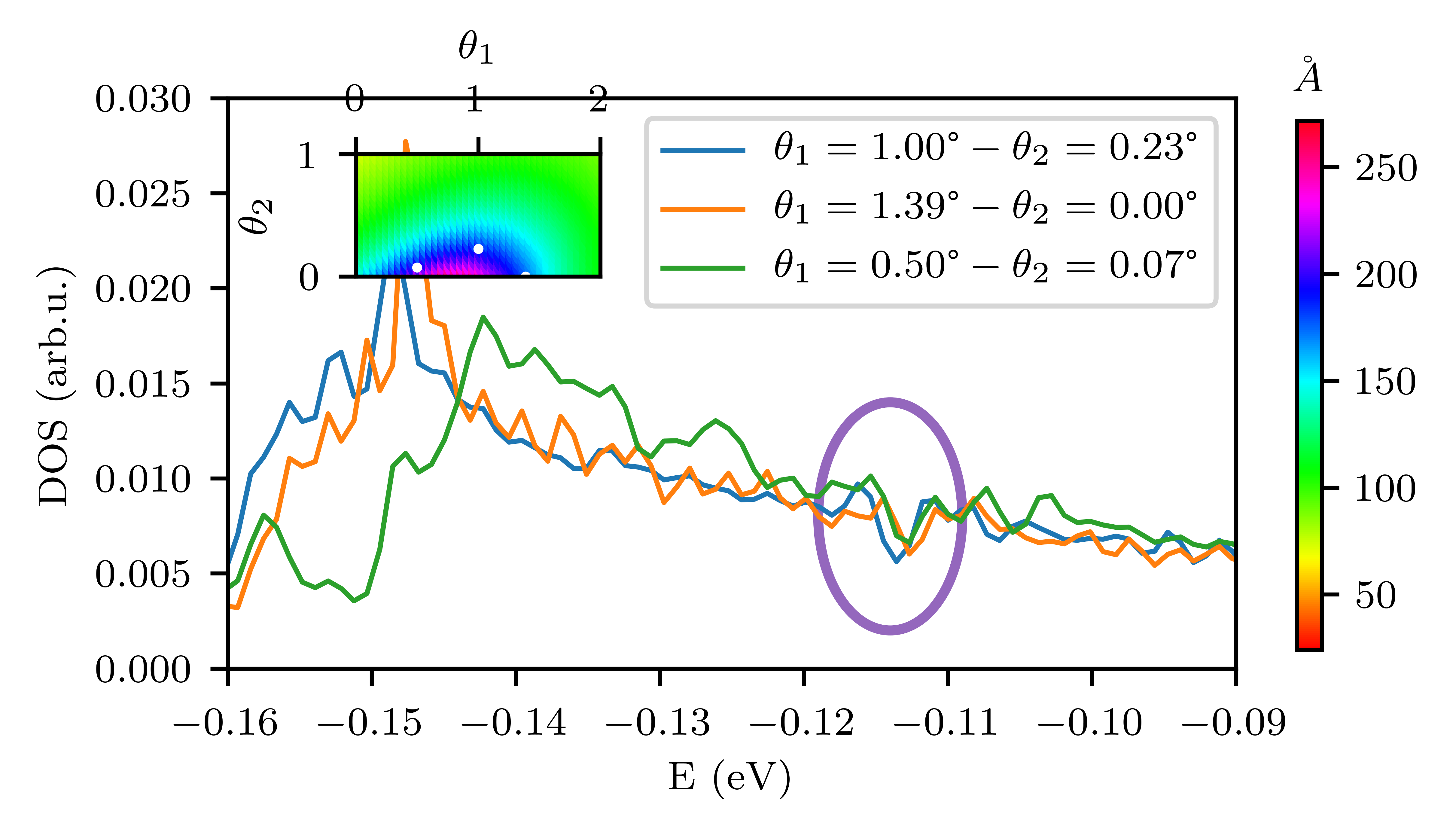}
\caption{\label{effectOfTheta} 
(Color online) Super-moire features in the DOS for $(p,q,r,s) = (1,1,1,0)$ circled in the figure obtained for 
several equal SM periods $\widetilde{\lambda}_{1110}$ obtained from different combinations of $\theta_1$ and $\theta_2$. 
The inset gives a map of the SM periods where the white dots indicate the angles used to obtain the curves from the main figure panel.
We find that the strength of the observed dip in the DOS is similar regardless of $\Delta \phi$ configuring the SM as well as the moire angles and lengths. The smallest oscillations in the figure have numerical origin. }
\end{center}
\end{figure}

So far we have shown that modifications in the electronic structure appear at energies that are 
related with the SM periods, yet these are in turn sensitive to the period and orientation of each moire pattern. 
In order to illustrate the sensitivity of SM effects and therefore of moire periods and orientations $\Delta \phi$ we 
show in Fig.~\ref{strongestfeatures} and Fig.~\ref{effectOfTheta} additional DOS for systems 
showing unequal twist angles between bottom and top BN layers. 
For Fig.~\ref{strongestfeatures} we have obtained the DOS maps for different sets of fixed bottom BN layer 
twist angles of $\theta_1 = 0.3\degree$, $0.6\degree$, $0.8\degree$, 
allowing for a variable top BN layer twist angle $\theta_2$.
These calculations confirm that the traces of the SM features in the DOS colormap closely follow the 
single major parameter $\widetilde{\lambda}_{pqrs}$, namely the SM period associated to given 
moire reciprocal lattice vectors and are insensitive to $\Delta \phi$. 
%
%
%
%
%
%
Specifically for fixed $\theta_1 = 0.6\degree$, when $\theta_2$ changes from $0$ to $1\degree$, 
$\Delta \phi$ varies continuously from about $65\degree$ down to about $-10\degree$ back 
up to about $20\degree$. Despite this wide changes in the moire twist angles we see a relative stability 
in the strength of the DOS features appearing on the hole side near the secondary Dirac point energies.
As $\theta_2 \rightarrow \theta_1$ the multiple supermoire features 
evolve into one single point as discussed earlier for commensurate systems. 
We further show in Fig.~\ref{strongestfeatures} relevant departures 
of the SM features on the DOS predicted from a Dirac dispersion through Eq.~(\ref{moireEnergy}) 
and the actual calculations where the features are pushed to slightly lower energies. 
This behavior can be fitted using a reduced Fermi velocity in the regime where the supermoire 
features are modifying the DOS over a larger part of the spectrum, e.g., in the middle panel, $-0.3$ to $-0.1$ eV for 
$\theta_1 = 0.6\degree$ and $\theta_2 = 0.0\degree$. 
We illustrate through black, white and red dashed lines the energies of the SM features from Dirac models using
$\upsilon_{\rm F} = 1.05 \upsilon_{{\rm F}_0}$, $1.01 \upsilon_{{\rm F}_0}$ and $0.93 \upsilon_{{\rm F}_0}$, respectively
where $\upsilon_{{\rm F}_0}$ is the Fermi velocity of graphene used in our simulations. 
The effective Fermi velocities of the Dirac model that fit best the SM features based on Eq.~(\ref{moireEnergy})
depend continuously on the $\theta_1$ and $\theta_2$ twist angles of the BN layers and the range of energies explored,
while it is expected that these SM features will be predicted more accurately by feeding the reference electronic structure 
bands from the G/BN moire bands corresponding of each BN interface rather than the bare Dirac cone dispersion.
We further confirm the relevance of the SM length $\widetilde{\lambda}_{pqrs}$ as the main parameter defining the energy 
scale of the SM features by comparing the DOS maps for three different cases that predict a $(p,q,r,s) = (1,1,1,0)$ SM 
feature at the same energy circled in purple in Fig.~\ref{effectOfTheta}. 
These three different cases have different moire lengths and different values of $\Delta \phi$, with the green line corresponding to a nearly 
perfectly aligned system, while the two others correspond to moire patterns that are not aligned). Yet we observe that the strength of the SM 
feature is nearly the same for the three systems, thus illustrating the relative insensitivity to misalignment.

\section{Conclusion}

In this manuscript we have examined the effects of double moire patterns created by 
two hexagonal boron nitride layers encapsulating single layer graphene. 
While the resulting electronic structure may at first order of approximation be explained 
from a linear superposition of each moire pattern, 
strong interference effects between the two moires are found when both BN layers are nearly aligned with graphene.
As a rule of thumb we can expect that stronger supermoire features will show up at energies 
closer to the charge neutrality point where the density of states
are low, \textit{i.e.} in the limit of $\widetilde{\lambda}_{\rm SM} \rightarrow \infty$ or 
conversely when the supermoire Brillouin zone boundary vector reaches values close to zero ${ k}_{\widetilde{M}} \rightarrow 0$. 
Near charge neutrality we can naturally expect enhancements in the band gaps when the
bottom and top BN layers reinforce the features reinforce each other. 
At the same time, the most prominent supermoire features away from charge neutrality are shown to appear at energies  
related with the position of the secondary Dirac point energies on the hole side in single graphene on boron nitride moire 
where a strong suppression in the density of states and a band gap is predicted near $-0.2$~eV for aligned orientations.
The most important interference effects between these secondary Dirac point features take place
for perfectly aligned moire patterns that we labeled as $C_0^{(\prime)}$, $C_1^{(\prime)}$ and $C_2^{(\prime)}$ phases
classified according to the relative rotation angle of the BN sheet with respect to graphene
where we find large DOS depressions leading to actual gaps and pseudo-gaps on the hole side 
of up to $\sim$0.5~eV in energy width at $-0.2$~eV and $-0.35$~eV, 
due to the convergence of a large number of supermoire features into a single 
energy value at the commensuration angles. 
Although the physics in the commensurate regime are rather sensitive to misalignment of the moire patterns 
it may be feasible to realize the commensurate double moires in actual 
experiments if we consider that locking them macroscopically can be energetically more stable than maintaining 
the misaligned double moire structures. 
Away from the commensuration angles between the moires we find that the supermoire length 
associated to specific moire reciprocal lattice vector combinations rather than their orientation 
is the main factor that determines the energy positions where the moire band features appear. 
We have also identified supermoire twist angles of moire quasi-crystals 
for $\Delta \phi =30\degree$ with dodecahedral symmetries that show van Hove singularities and  
multiple gaps and oscillations in the density of states that deserve further studies.
In summary we have shown that in graphene encapsulated by boron nitride systems the strongest double moire interference 
effects are expected commensurate double moire structres that can give rise to large modifications in the electronic structure 
manifested in the formation of new band gaps and van Hove singularities. 
These observations indicate optimistic prospects of achieving 
new electronic structures that hosts strong correlation effects by means of double moire interference
in van der Waals heterostructures.

%
  
\ack
We gratefully acknowledge computational resources from Texas Advanced Computing Centre (TACC) 
and KISTI through grant KSC-2018-CHA-0077. We thank Mr. Junhyuk Park for assistance with some figures.
N. L. acknowledges support from the National Research Foundation (NRF) of Korea through grant number 
NRF-2018R1C1B6004437 and the Korea Research Fellowship Program through the NRF funded by the Ministry and Science and 
ICT 610 (KRF-2016H1D3A1023826).
J. J. acknowledges support from Samsung Science and Technology Foundation under project no. SSTF-BA1802-06.

\newpage

\bibliographystyle{unsrt}
\bibliography{bib.bib}


\end{document}